\documentclass{mem}
\usepackage{natbib}
\usepackage[varg]{txfonts}
\usepackage{balance}
\usepackage{graphicx}
\usepackage[a4paper,breaklinks,dvipdfm]{hyperref}
\idline{75}{282}
\newcommand{\etal}{et al.}
\begin{document}

\title{$^6$Li detection in metal-poor stars:
can 3D model atmospheres solve the second lithium problem?}


\author{
M. \,Steffen\inst{1,2}
\and \,R.~Cayrel\inst{2}
\and \,E.~Caffau\inst{3,2}
\and \,P.~Bonifacio\inst{2}
\and \,H.-G.~Ludwig\inst{3,2}
\and \,M.~Spite\inst{2}
          }
 
\institute{
Leibniz-Institut f\"ur Astrophysik Potsdam (AIP), An der Sternwarte 16,
14482 Potsdam, Germany,
\email{msteffen@aip.de}
\and
GEPI Observatoire de Paris, CNRS, Universit´e Paris Diderot, 
F-92195 Meudon Cedex, France
\and
Zentrum f\"ur Astronomie der Universit\"at Heidelberg, Landessternwarte, 
K\"onigstuhl 12, 69117 Heidelberg, Germany
}

\authorrunning{M.Steffen \etal}

\titlerunning{$^6$Li detection in metal-poor stars}

\abstract{
The presence of $^6$Li in the atmospheres of metal-poor halo stars is usually
inferred from the detection of a subtle extra depression in the red wing of
the $^7$Li doublet line at 670.8 nm. However, as pointed out recently by
\cite{Cayrel2007}, the intrinsic line asymmetry caused by convective flows in
the photospheres of cool stars is almost indistinguishable from the asymmetry
produced by a weak $^6$Li blend on a (presumed) symmetric $^7$Li
profile. Previous determinations of the $^6$Li/ $^7$Li isotopic ratio based on
1D model atmospheres, ignoring the convection-induced line asymmetry, must
therefore be considered as upper limits.
By comparing synthetic 1D LTE and 3D non-LTE line profiles of the
\ion{Li}{i}\,$670.8$~nm feature, we quantify the differential effect of the
convective line asymmetry on the derived $^6$Li abundance as a function of
effective temperature, gravity, and metallicity.  As expected, we find that
the asymmetry effect systematically reduces the resulting $^6$Li/$^7$Li
ratios. Depending on the stellar parameters, the 3D-1D offset in $^6$Li/$^7$Li
ranges between $-0.005$ and $-0.020$. When this purely theoretical correction
is taken into account for the \cite{A2006} sample of stars, the number of
significant $^6$Li detections decreases from 9 to 5 (2$\sigma$ criterion), or
from 5 to 2 (3$\sigma$ criterion).

We also present preliminary results of a re-analysis of
high-resolution, high S/N spectra of individual metal-poor turn-off stars, to
see whether the \emph{second Lithium problem} actually disappears when
accounting properly for convection and non-LTE line formation in 3D stellar
atmospheres. Out of $8$ stars, HD\,84937 seems to be the only significant
(2$\sigma$) detection of $^6$Li. In view of our results, the existence of a
$^6$Li plateau appears questionable.

\\[-4mm]
}

\maketitle{}

\section{Introduction}
A systematic analysis of high-dispersion, high signal-to-noise (S/N) UVES 
and Keck spectra of about 30 bright metal-poor stars by \cite{A2006} and 
\cite{A2008} (henceforth A06 and A08, respectively) resulted in the detection 
of $^6$Li (at the $2\, \sigma$ level) in about one third of these objects.  
The average $^6$Li/$^7$Li isotopic ratio in the stars in
which $^6$Li has been detected is about 4\% and is very similar in each of
these stars, defining a $^6$Li plateau at approximately 
$\log n(^6$Li$)/n($H$)+12 = 0.85$.  A convincing theoretical explanation
of this new $^6$Li plateau turned out to be problematic: the high abundance of
$^6$Li at the lowest metallicities cannot be explained by current models of
galactic cosmic-ray production, even if the depletion of $^6$Li during the 
pre-main-sequence phase is ignored \citep[see reviews by e.g.]
[and references therein]{Cayrel2008, Ch2008, Prantzos2010, Prantzos2012}.

A possible solution of this so-called \emph{second Lithium problem} 
was proposed by \cite{Cayrel2007}: noting that the spectroscopic signature of
the presence of $^6$Li in the atmospheres of metal-poor halo stars is just a
subtle extra depression in the red wing of the $^7$Li doublet, which is
difficult to measure even in spectra of the highest quality, they point out
that the intrinsic line asymmetry caused by convection in the photospheres of
cool stars closely mimics the presence of $^6$Li at the level of a few percent
if interpreted in terms of 1D, intrinsically symmetric blend components. As a
consequence, the $^6$Li abundance derived so far by using 1D model
atmospheres, ignoring the convective line asymmetry, are expected to be
systematically too high.

We quantify the theoretical effect of the convection-induced line
asymmetry on the resulting $^6$Li abundance by fitting a given 
synthetic profile both with a grid of 1D LTE and a grid of 3D non-LTE
line profiles. The 1D-3D difference of the required $^6$Li/$^7$Li isotopic 
ratio measures the expected systematic error in $^6$Li/$^7$Li inherent
to the standard 1D analysis. The synthetic spectra used for this
differential comparison are based on a set of 3D hydrodynamical model 
atmospheres computed with the CO$^5$BOLD code \citep{F2002,F2012}
and a corresponding set of fully compatible 1D mixing-length models,
respectively, as outlined in {Sect.\,\ref{s:3Dmodels}}. 
The results of this investigation may be represented 
by an analytical approximation, giving the 3D correction of $^6$Li/$^7$Li
as a function of effective temperature, gravity, and metallicity, for a 
parameter range that covers the stars of the A06 sample
({Sect.\,\ref{s:3Dbias}}).
We also discuss the potential advantages and disadvantages
of using so called 'calibration lines' for fixing the residual line 
broadening, and show that the usage of such 'calibration lines' is 
potentially dangerous, because the inferred broadening parameter shows 
considerable line-to-line variations. Depending on the choice of these
lines, the resulting $^6$Li abundance may be systematically biased
({Sect.\,\ref{s:VBRcalibration})}.

A careful reanalysis of individual objects is under 
way to see whether the \emph{second Lithium problem} can be resolved
when accounting properly for convection and non-LTE line formation in 3D 
stellar atmospheres. In {Sect.\,\ref{s:8stars}} we present a preliminary 
analysis of eight individual metal-poor turn-off stars for which sufficiently
high-resolution, high S/N spectra are at our disposal, fitting the observed 
\ion{Li}{i}\,670.8~nm feature with both 1D LTE and 3D non-LTE synthetic line 
profiles. As expected, the 3D analysis gives a systematically lower
$^6$Li/$^7$Li ratio by roughly $0.01$ with respect to the 1D result. 
In most cases, we find that the detection of $^6$Li is not significant at 
the $2\sigma$ level.

\section{3D model atmospheres and spectrum synthesis}
\label{s:3Dmodels}
The hydrodynamical atmospheres used in the present study are part of the
CIFIST 3D model atmosphere grid \citep{L2009}. They have
been obtained from realistic numerical simulations with the CO$^5$BOLD 
code\footnote{\url{http://www.astro.uu.se/~bf/co5bold_main.html}}
\citep{F2002,F2012} which solves the time-dependent equations of compressible 
hydrodynamics in a constant gravity field together with the equations of 
non-local, frequency-dependent radiative transfer in a Cartesian box 
representative of a volume located at the stellar surface. The computational 
domain is periodic in $x$ and $y$ direction, has open top and bottom 
boundaries, and is resolved by typically 140$\times$140$\times$150 grid
cells. The vertical optical depth of the box varies typically from 
$\log \tau_{\rm Ross}$$\approx$$-7.0$ (top) to 
$\log \tau_{\rm Ross}$$\approx$$+7.5$ (bottom). 
Further information about the grid of models used in the present
study is compiled in Table\,\ref{tab1}. As indicated in columns (2) -- (4),
the grid covers the range $5900$~K $\le T_{\rm eff}$ $\le 6500$~K, 
$3.5$ $\le \log g$ $\le 4.5$, $-3.0$ $\le [\mathrm{Fe/H}]$ $\le 0.0$.
Radiative transfer is solved in $5$ or $6$ opacity bins (col.\ 5). 
Each of the models is represented by typically $20$ snapshots chosen 
from the full time sequence of the corresponding simulation (col.\ 6).
 
These representative snapshots are processed by the non-LTE code NLTE3D that 
solves the statistical equilibrium equations for a 17 level lithium atom with 
34 line transitions, fully taking into account the 3D thermal structure of
the respective model atmosphere (but neglecting the differential Doppler
shifts implied by the hydrodynamical velocity field). 
The photo-ionizing radiation field is
computed at $704$ frequency points between $\lambda\,925$ and 32\,407~\AA,
using the opacity distribution functions of \cite{CK2004} to allow for
metallicity-dependent line-blanketing, including the H\,I--H$^+$ and 
H\,I--H\,I quasi-molecular absorption near $\lambda\,1400$ and $1600$~\AA, 
respectively. Collisional ionization by neutral hydrogen via the charge 
transfer reaction 
H($1s$) + Li($n\ell$) $\leftrightarrow$ Li$^+$($1s^2$) + H$^-$ is 
treated according to \cite{barklem2003}. More details are given in 
\cite{S2010}. Finally, 3D non-LTE synthetic line profiles of the \ion{Li}{i} 
$\lambda\,670.8$~nm doublet are computed with the line formation code 
Linfor3D\footnote{\url{http://www.aip.de/~mst/linfor3D_main.html}}, using
the departure coefficients $b_i$\,=\,$n_i({\rm NLTE})/n_i({\rm LTE})$
provided by NLTE3D for each level $i$ of the lithium model atom as a function
of geometrical position within the 3D model atmospheres. At this stage, the
3D hydrodynamical velocity field provides the differential line shifts along
each line of sight. As demonstrated in
\cite{steffen2010b}, non-LTE effects are very important for the 3D models of
the metal-poor dwarfs considered here: they strongly reduce the height range 
of line formation such that the 3D non-LTE equivalent width is smaller by 
roughly a factor 2 compared to 3D LTE. Ironically, the line strength predicted
by standard 1D mixing-length models in LTE are close to the results obtained
from elaborate 3D non-LTE calculations. Nevertheless, when it comes to the
determination of the $^6$Li/$^7$Li isotopic ratio, it is important to account
for the convective line asymmetry, which requires the full 3D non-LTE line
formation calculations.

\begin{figure*}[htb]
\resizebox{\hsize}{!}{\includegraphics[bb=40 28 580 380,clip=true]
{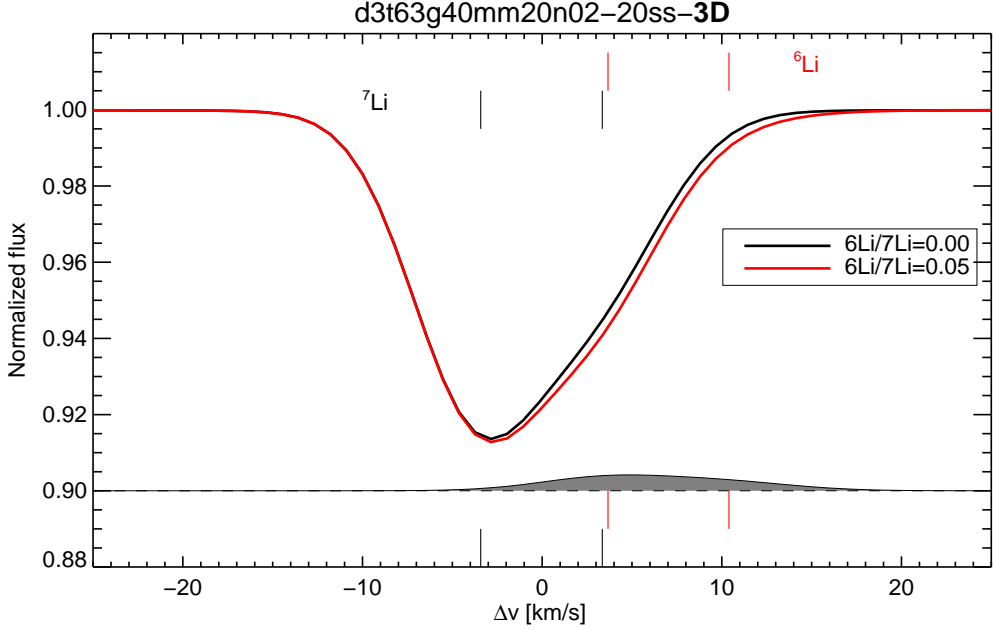}}
\caption{\footnotesize
Example of a 3D non-LTE synthetic profile of the \ion{Li}{i} $670.8$~nm 
feature, for stellar parameters $T_{\rm eff}=6300$~K, $\log g=4.0$, 
[Fe/H]=-2. Increasing the $^6$Li content from $0$ to $5$\% leads to an extra
depression of the red wing by about $0.5$\% (indicated by the gray area), 
leaving the blue wing unaffected. In order to detect such a small effect, 
observed spectra are required to have an S/N in excess of $\approx 400$.
}
\label{fig01}
\end{figure*}
\begin{table*}[bt]
  \begin{center}
  \caption{List of 3D models used in the present study. Columns (2)-(6) give 
           effective temperature, surface gravity, metallicity, number of 
           opacity bins used in the radiation hydrodynamics simulation, and 
           number of snapshots selected for spectrum synthesis. The equivalent 
           width of the synthetic 3D non-LTE $^7$Li doublet at 
           $\lambda\,670.8$~nm, assuming $A$(Li)=2.2 and no $^6$Li, 
           is given in col.\ (7). Columns (8) and (9) show 
           $\Delta q_{\rm (A)}$(Li) and $\Delta q_{\rm (B)}$(Li), the 3D 
           non-LTE correction of the $^6$Li/$^7$Li isotopic ratio inferred 
           from fitting the unprocessed (no rotational and instrumental 
           broadening) and the broadened ($v\,\sin i = 1.5$~km/s, 
           $FWHM=2.5$~km/s) 3D non-LTE line profile, respectively, with 
           a grid of 1D LTE profiles.}
  \label{tab1}
 {\small
  \begin{tabular}{ccccccccc}\hline
{Model}               & {$T_{\rm eff}^{~~1)}$} & {$\log g$}    & 
{[Fe/H]}              & {Nbin}                & {Nsnap}     &
{W$^{ 2)}$}            & \multicolumn{2}{c}{$\Delta q$(Li)=$q^\ast_{\rm(1D\,LTE)}$(Li)} \\
{}                        & {[K]}                 &                 & 
                          &                           &                 & 
{[m\AA]}                  & \multicolumn{2}{c}{[$\times 100$]}       \\
& & & & & & & \hspace{2.5mm} {(A)} \hspace{2.5mm}& {(B)} \\
\hline
d3t59g35mm30n01& $5873$ & $3.5$ & -3.0 & $  6$ & 20 & 41.7 & {2.13}& {1.99} \\
d3t59g40mm30n02& $5846$ & $4.0$ & -3.0 & $  6$ & 20 & 44.9 & {0.88}& {0.86} \\
d3t59g45mm30n01& $5924$ & $4.5$ & -3.0 & $  6$ & 19 & 39.9 & {0.63}& {0.61} \\
d3t63g35mm30n01& $6306$ & $3.5$ & -3.0 & $  6$ & 20 & 21.7 & {3.44}& {3.12} \\
d3t63g40mm30n01& $6269$ & $4.0$ & -3.0 & $  6$ & 20 & 23.9 & {1.51}& {1.38} \\
d3t63g45mm30n01& $6272$ & $4.5$ & -3.0 & $  6$ & 18 & 24.3 & {0.85}& {0.80} \\
d3t65g40mm30n01& $6408$ & $4.0$ & -3.0 & $  6$ & 20 & 20.0 & {1.70}& {1.56} \\
d3t65g45mm30n01& $6556$ & $4.5$ & -3.0 & $  6$ & 12 & 16.4 & {1.25}& {1.16} \\
\hline                                                                          
d3t59g35mm20n01& $5861$ & $3.5$ & -2.0 & $  6$ & 20 & 42.8 & {1.88}& {1.76} \\
d3t59g40mm20n02& $5856$ & $4.0$ & -2.0 & $  6$ & 20 & 45.2 & {0.73}& {0.69} \\
d3t59g45mm20n01& $5923$ & $4.5$ & -2.0 & $  6$ & 18 & 42.3 & {0.45}& {0.41} \\
d3t63g35mm20n01& $6287$ & $3.5$ & -2.0 & $  6$ & 20 & 22.1 & {4.04}& {3.75} \\
d3t63g40mm20n01& $6278$ & $4.0$ & -2.0 & $  6$ & 16 & 23.7 & {1.48}& {1.35} \\
d3t63g45mm20n01& $6323$ & $4.5$ & -2.0 & $  6$ & 19 & 23.0 & {0.94}& {0.88} \\
d3t65g40mm20n01& $6534$ & $4.0$ & -2.0 & $  6$ & 19 & 16.2 & {2.22}& {2.01} \\
d3t65g45mm20n01& $6533$ & $4.5$ & -2.0 & $  6$ & 19 & 17.0 & {1.21}& {1.12} \\
\hline                                                                          
d3t59g35mm10n02& $5890$ & $3.5$ & -1.0 & $  6$ & 20 & 38.0 & {2.91}& {2.71} \\
d3t59g40mm10n02& $5850$ & $4.0$ & -1.0 & $  6$ & 20 & 41.5 & {1.45}& {1.36} \\
d3t59g45mm10n01& $5923$ & $4.5$ & -1.0 & $  6$ & 08 & 38.2 & {0.83}& {0.76} \\
d3t63g35mm10n01& $6210$ & $3.5$ & -1.0 & $  6$ & 20 & 23.0 & {4.57}& {4.18} \\
d3t63g40mm10n01& $6261$ & $4.0$ & -1.0 & $  6$ & 20 & 22.0 & {2.33}& {2.15} \\
d3t63g45mm10n01& $6238$ & $4.5$ & -1.0 & $  6$ & 20 & 23.4 & {1.23}& {1.14} \\
d3t65g40mm10n01& $6503$ & $4.0$ & -1.0 & $  6$ & 20 & 15.5 & {3.14}& {2.82} \\
d3t65g45mm10n01& $6456$ & $4.5$ & -1.0 & $  6$ & 19 & 17.1 & {1.44}& {1.33} \\
\hline                                                                          
d3t59g35mm00n01& $5884$ & $3.5$ &  0.0 & $  5$ & 20 & 33.0 & {3.37}& {3.06} \\
d3t59g40mm00n01& $5928$ & $4.0$ &  0.0 & $  5$ & 18 & 31.1 & {1.83}& {1.70} \\
d3t59g45mm00n01& $5865$ & $4.5$ &  0.0 & $  5$ & 19 & 34.7 & {1.15}& {1.06} \\
d3t63g40mm00n01& $6229$ & $4.0$ &  0.0 & $  5$ & 20 & 19.1 & {2.66}& {2.43} \\
d3t63g45mm00n01& $6233$ & $4.5$ &  0.0 & $  5$ & 20 & 19.3 & {1.49}& {1.47} \\
d3t65g40mm00n01& $6484$ & $4.0$ &  0.0 & $  5$ & 20 & 12.8 & {3.88}& {3.55} \\
d3t65g45mm00n01& $6456$ & $4.5$ &  0.0 & $  5$ & 20 & 13.8 & {1.75}& {1.61} \\
\hline
  \end{tabular}
  }
 \end{center}
\vspace{1mm}
 \scriptsize{
 {\it Notes:}
  $^{1)}$ averaged over selected  snapshots; $^{2)}$ averaged over selected 3D non-LTE 
  spectra}
\end{table*}

\section{Theoretical 3D non-LTE correction of 1D LTE 
$^6$Li/$^7$Li determinations}
\label{s:3Dbias}
As outlined above, the $^6$Li abundance is systematically overestimated if
the intrinsic asymmetry of the $^7$Li line components is ignored. To quantify 
this bias theoretically, we rely on synthetic spectra. The idea is as follows: 
we represent the observation by the synthetic 3D non-LTE line profile of the 
$^7$Li line blend, computed with zero $^6$Li content. The non-thermal line
broadening is naturally provided by the 3D hydrodynamical velocity field, 
which replaces the classical concept of micro- and macroturbulence, and also
gives rise to a convective blue-shift and an intrinsic line asymmetry. 
Optionally, the line blend is subsequently broadened by rotation ($v\,\sin i$) 
and a Gaussian instrumental profile ($FWHM$ denoting its full width at half 
maximum). Now this realistic proxy of the $^7$Li line blend is fitted by a 
grid of 'classical' 1D LTE synthetic line profiles. The 1D LTE line profiles 
are computed from so-called LHD models, 1D mixing-length model atmospheres 
that have the same stellar parameters and employ the same microphysics and
radiative transfer scheme as the corresponding 3D model, assuming a
mixing-length parameter of $\alpha_{\rm MLT}=0.5$, and a depth-independent
microturbulence of $\xi_{\rm mic}=1.5$~km/s.

\begin{figure*}[htb]
\mbox{\includegraphics[width=0.47\hsize,bb=36 32 560 370,clip=true]
{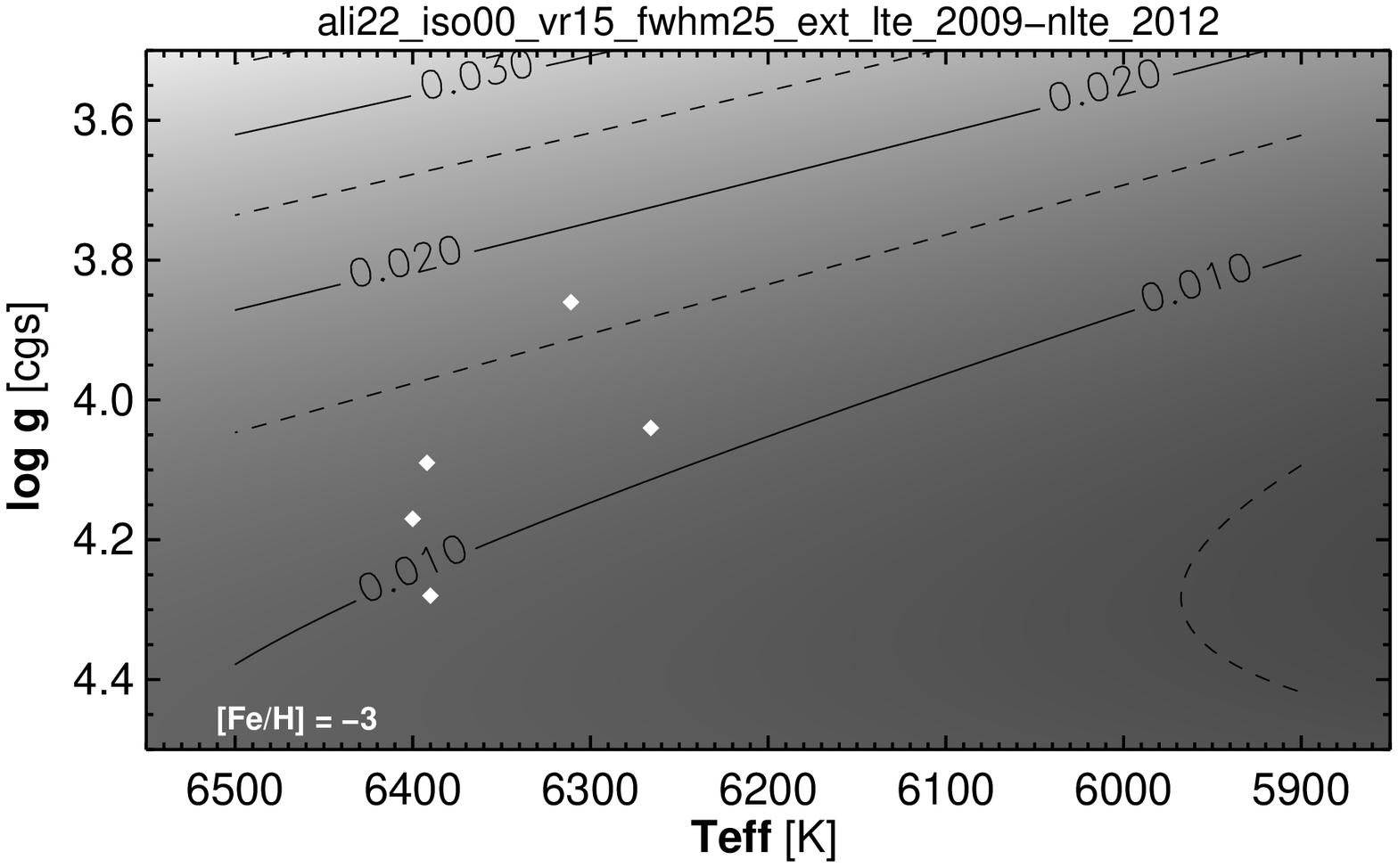}}
\mbox{\includegraphics[width=0.47\hsize,bb=36 32 560 370,clip=true]
{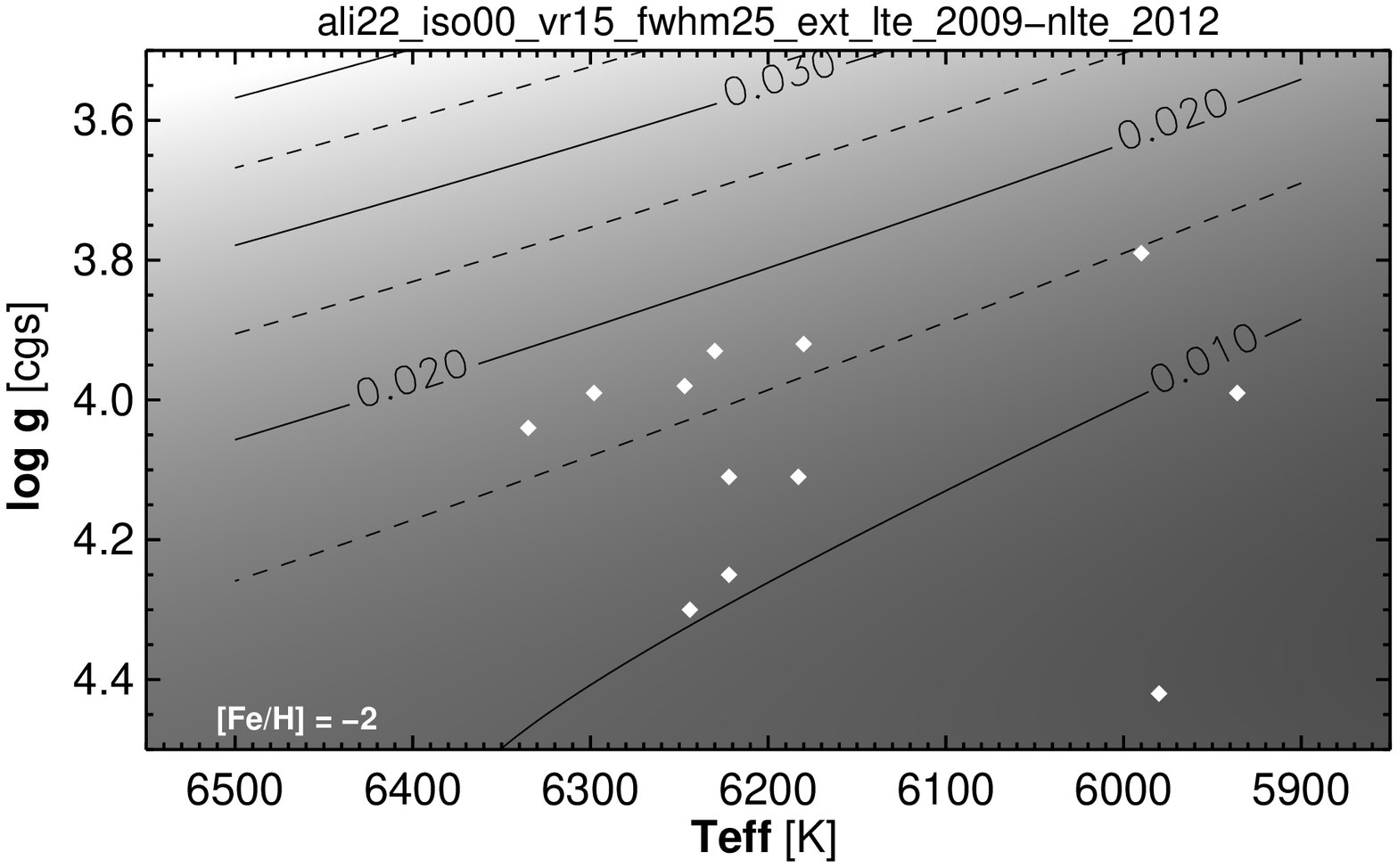}}\\[2mm]
\mbox{\includegraphics[width=0.47\hsize,bb=36 32 560 345,clip=true]
{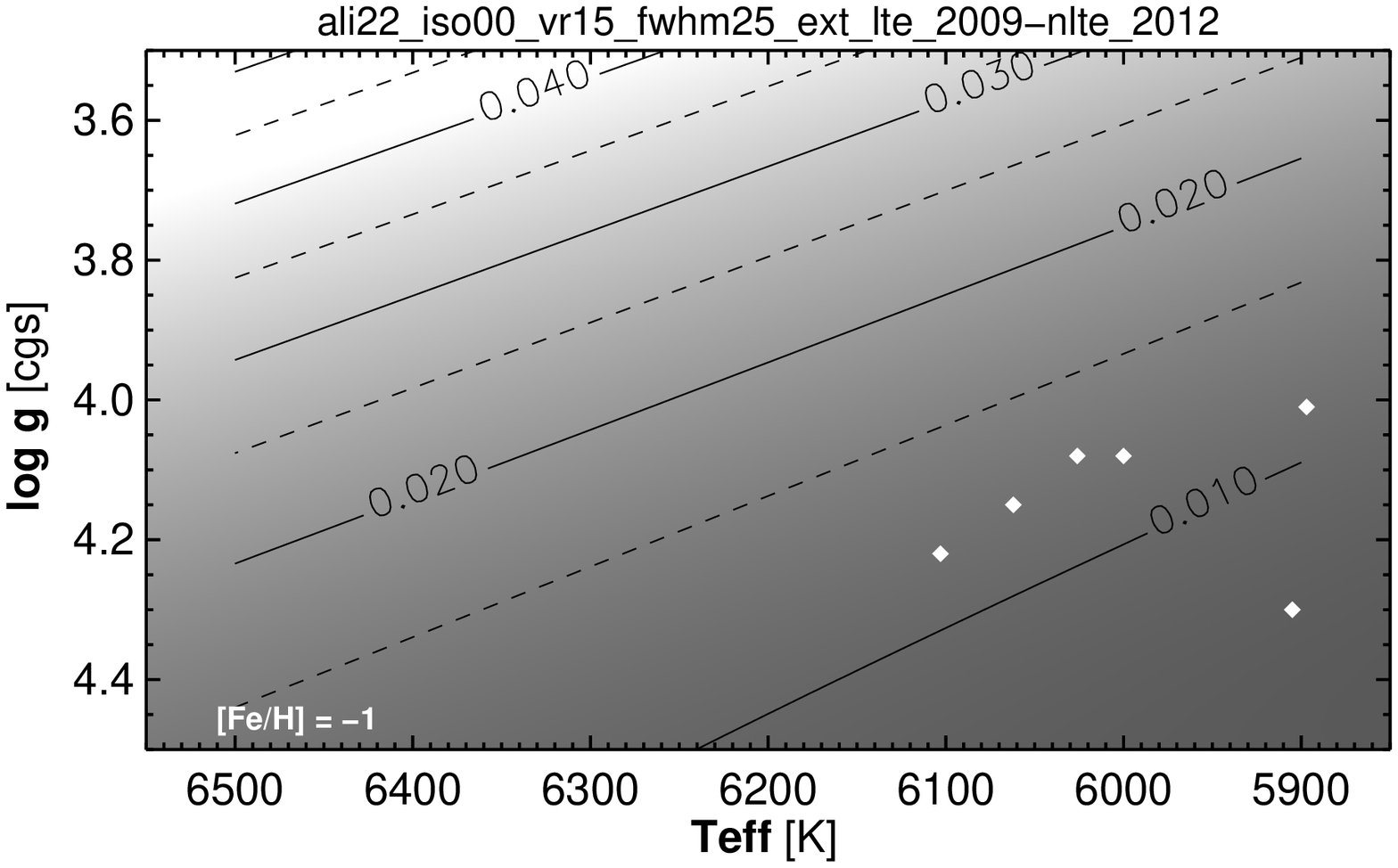}}
\mbox{\includegraphics[width=0.47\hsize,bb=36 32 560 345,clip=true]
{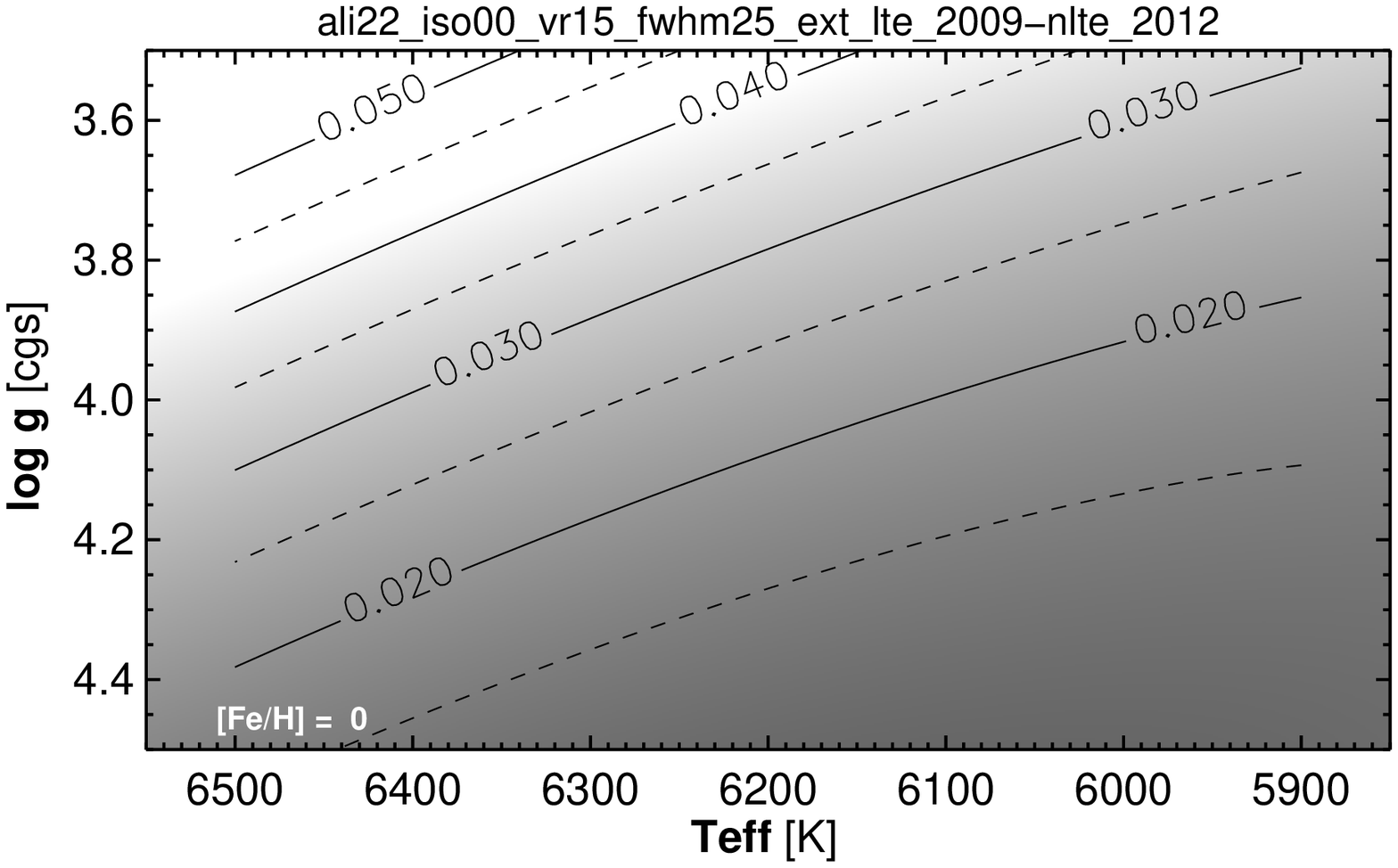}}
\caption{\footnotesize
Contours of $\Delta q_{\rm (B)}$(Li) in the $T_{\rm eff}$ -- $\log g$ plane
at constant metallicity [Fe/H] = $-3$, $-2$, $-1$, and $0$, respectively
(from top left to bottom right) according to the approximation given by 
Eq.\,(\ref{eqn1}) and Table\,\ref{tab2}. White symbols mark the positions 
of the stars from the A06 sample with Fe/H] $< -2.5$ (top left),
$-2.5<$ [Fe/H] $<-1.5$ (top right), and $-1.5<$ [Fe/H] $<-0.5$ 
(bottom left).
}
\label{fig02}
\end{figure*}

Five parameters are varied independently until the best fit is
found ($\chi^2$ minimization):
in addition to the total $^6$Li+$^7$Li abundance, $A$(Li), and the isotopic 
ratio, $q$(Li)=$^6$Li/$^7$Li, which control line strength and line asymmetry, 
respectively, we also allow for a residual line broadening described by a 
Gaussian kernel with half-width $V_{\rm BR}$, a global line shift, $\Delta v$,
and a global scaling of the continuum level, $S_{\rm c}$. For technical 
reasons, the rotational line broadening is not treated as a fitting parameter; 
it is fixed to the value used in the 3D spectrum synthesis. 

Note that, although the five fitting parameters are correlated, 
each one has a distinctly different effect on the line profile. 
The broadening $V_{\rm BR}$ is fixed by the slope of the blue wing,
which is not affected by the $^6$Li content, (see Fig.\,\ref{fig01}),
while $q$(Li) is fixed by the shape of the red wing.
This allows to obtain unambiguously a unique solution
although a given half width of the line blend can be obtained by 
different combinations of $q$(Li) and $V_{\rm BR}$, but with higher
$\chi^2$ values than the best solution.

Finally, the 3D non-LTE correction of the $^6$Li/$^7$Li isotopic ratio is 
defined as the difference between the values of $q$(Li) that provide 
the best fit ($^\ast$) to the given profile using 1D LTE and 3D non-LTE 
profiles, respectively:
$\Delta q$(Li) = $q^\ast_{\rm (1D\,LTE)}$(Li) - $q^\ast_{\rm (3D\,NLTE)}$(Li)%
\footnote{$q^\ast_{\rm (3D\,NLTE)}$(Li) is zero by construction here}. 
This differential correction is meant to be 
\emph{subtracted} from the $^6$Li/$^7$Li isotopic ratio derived from the 
1D LTE analysis to correct for the bias introduced by neglecting the 
intrinsic line asymmetry: 
$q_{\rm (3D\,NLTE)}$(Li) = $q_{\rm (1D\,LTE)}$(Li) - $\Delta q$(Li). 
Note that this correction procedure properly accounts for radiative transfer 
in the lines, including saturation effects.

We have determined $\Delta q$(Li) according to the method outlined above
for each of the 3D model atmospheres in our grid, considering two cases:
$\Delta q_{\rm (A)}$(Li) and $\Delta q_{\rm (B)}$(Li) are the result of
fitting the unprocessed
($v\,\sin i = 0.0$~km/s, $FWHM=0.0$~km/s) and the broadened
($v\,\sin i = 1.5$~km/s, $FWHM=2.5$~km/s) synthetic line profile,
respectively. Case (B) is the more realistic one, while case (A) has 
been included only to show how the amount of instrumental broadening 
is affecting the amplitude of $\Delta q$(Li). Note also that the 3D 
corrections given in previous studies by 
\citep{steffen2010a,steffen2010b} correspond to case (A). 

The results of this computationally very expensive procedure are given 
in cols.\ (8) and (9) of Table\,\ref{tab1}. They may be aproximated 
fairly well by the following polyonmial expression:
\begin{eqnarray}
\Delta q_{\rm (B)}\mathrm{(Li)}\{T_{\rm eff},\log g,\mathrm{[Fe/H]}\} =
a_{00} + a_{01}\,Z &&+ \nonumber \\
\left(a_{10} + a_{11}\,Z\right)\,Y    + 
\left(a_{20} + a_{21}\,Z\right)\,Y^2  &&+ \nonumber \\
\left(a_{30} + a_{31}\,Z\right)\,X    +
\left(a_{40} + a_{41}\,Z\right)\,X\,Y &&+ \nonumber \\
\left(a_{50} + a_{51}\,Z + a_{52}\,Z^2\right)\,X^2&&
\label{eqn1}
\end{eqnarray}
where $X$$\equiv$$(T_{\rm eff}-6300)/1000$, $Y$$\equiv$$\log g - 4.0$, and 
$Z$$\equiv$[Fe/H]+2. The coefficients $a_{ij}$ are listed in 
Table\,\ref{tab2}. Contour plots of this polynomial approximation are 
shown in Fig.\,\ref{fig02}.

\begin{table}
\caption{Coefficients $a_{\rm ij}$ representing $\Delta q_{\rm (B)}$(Li) as a 
function of $T_{\rm eff}$, $\log g$, and [Fe/H] according to Eq.\,(\ref{eqn1}).}
\label{tab2}
\begin{center}
\begin{tabular}{lrrr}
\hline
\\
 j  & $a_{\rm 0j}\times 100$ & $a_{\rm 1j}\times 100$ & 
      $a_{\rm 2j}\times 100$ \\
\hline
\\
0 &   1.6984 &  0.4311 &  0.0000 \\
1 &  -2.6658 & -0.4547 &  0.0000 \\
2 &   2.3370 & -0.3211 &  0.0000 \\
3 &   2.3490 &  0.6300 &  0.0000 \\
4 &  -2.7366 & -0.6089 &  0.0000 \\
5 &   0.1786 &  0.5657 &  0.5385 \\
\\
\hline
\end{tabular}
\end{center}
\end{table}

At given metallicity, the corrections are largest for low gravity and high
effective temperature. They increase towards higher metallicity.
Note that the corrections are strictly valid only for a Li abundance
of $A$(Li)=2.2. The results for [Fe/H]=0.0 are most uncertain, since 
they are based on fewer data points.

\begin{figure}[htb]
\mbox{\includegraphics[width=\hsize,bb=40 28 580 400,clip=true]
{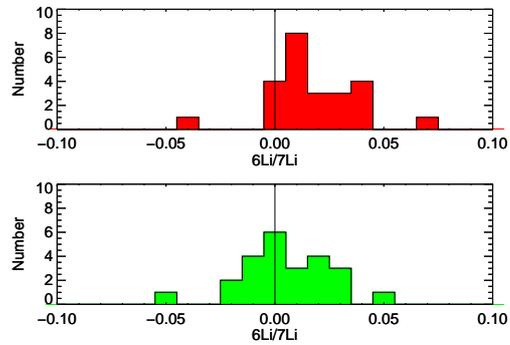}}
\caption{\footnotesize
Histogram of the $^6$Li/$^7$Li determinations by A06 before
(top) and after (bottom) application of the 3D correction
$\Delta q_{\rm (B)}$(Li).
}
\label{fig03}
\end{figure}

The analysis of A06 utilizes 1D LTE profiles computed from
MARCS model atmospheres. Hence, the correction $\Delta q_{(B)}$(Li) 
should be applied to their $^6$Li/$^7$Li isotopic ratios. The resulting 
downward corrections are typically in the range $1\% < \Delta q_{(B)}$(Li) 
$< 2\%$ for the stars of their sample (cf.\ Fig.\,\ref{fig02}).
After subtracting the individual $\Delta q_{(B)}$(Li) for each of these
stars, interpolated via Eq.\,(\ref{eqn1}) to their $T_{\rm eff}$, 
$\log g$, and [Fe/H], the mean $^6$Li/$^7$Li isotopic ratio of the sample is
reduced from $0.0212$ to $0.0088$. The center of the distribution of 
$^6$Li/$^7$Li is shifted essentially to zero (see Fig.\,\ref{fig03}).

Based on the original $^6$Li/$^7$Li isotopic ratios and their $1\sigma$ 
error bars as determined by A06, the number of stars with a 
$^6$Li detection above the $2\sigma$ and $3\sigma$ level is $9$ and $5$, 
respectively, out of $24$. After correction, the number of $2\sigma$ and 
$3\sigma$ detections is reduced to $5$ and $2$, respectively. 
The $3\sigma$ detections after correction are {HD\,106038} and 
{G020$-$024}, the remaining $2\sigma$ detections are {HD\,102200}, 
{HD\,160617}, and {CD$-$30\,18140}. 

We note that HD\,106038 survives as a $3\sigma$ detection only because of 
its amazingly small error bar of $\pm 0.006$. Assuming a more typical error
of $\pm 0.013$, this object would not even qualify as a $2\sigma$ detection.
HD\,102200 seems to be a clear 2$\sigma$ detection ($q$(Li)~$=0.036 
\pm 0.013$), while CD$-$30\,18140 is a marginal case ($q$(Li)~$=0.029
\pm 0.013$). The remaining objects, G020-024 and HD\,160617, are discussed 
in more detail in Sect.\,\ref{s:8stars}, where their spectra are reanalyzed 
with 3D non-LTE line profiles; none of them provides convincing evidence for 
the presence of $^6$Li.

\section{Fixing the residual line broadening by 'calibration lines'}
\label{s:VBRcalibration}
So far we have derived all information from fitting the Li $670.8$~nm
feature, for which we have a 3D non-LTE line formation model.
In principle, the accuracy of the fitting procedure can be improved
by fixing the value of the residual line broadening, $V_{\rm BR}$,
from additional unblended 'calibration lines'. This approach has been 
adopted by A06 \citep[see also][henceforth L12]{L2012}. This fact 
complicates the comparison with our results, and the correction procedure
described in Sect.\,\ref{s:3Dbias} explains only part of the differences. 
We show in the following that 'calibration lines' may introduce additional
uncertainties, unless their 3D non-LTE line formation is fully understood.

In the framework of 1D models, $V_{\rm BR}$, represents the combined effect 
of macroturbulence and instrumental broadening (for given microturbulence 
and rotational velocity) and is expected to be independent of the spectral
line. In general, an average value of $V_{\rm BR}$ from several calibration
lines is then used for the analysis of the Li line, thus reducing the number
of free fitting parameters, and hence the formal errors of the fitting
results.

\begin{table}
\caption{List of calibration lines. Columns (2) to (4) show their
wavelength, excitation potential, and equivalent width. The residual 
line broadening derived from 1D LTE fitting and the resulting error 
of $^6$Li/$^7$Li is given in cols.\ (5) and (6). The last row refers
to the Li line; cols.\ (5) and (6) show the results of 3D non-LTE fitting 
without calibration lines.} 
\label{tab3}
\begin{center}
\begin{tabular}{lrrrrr}
\hline\noalign{\smallskip}
 Ion  & $\lambda$ & $E_{\rm i}$ & 
      $W_\lambda$ & $V_{\rm BR}$ & $\delta q^{(1D)}$ \\
      & [nm] & [eV] & [pm] & [kms] & [$\times 100$] \\
\hline
\\
\ion{Fe}{i} &   538.337  & 4.312 &  2.74 & 4.88 & {0.35} \\
\ion{Fe}{i} &   606.548  & 2.608 &  1.38 & 4.63 & {1.25} \\
\ion{Fe}{i} &   613.662  & 2.453 &  2.47 & 4.52 & {1.64} \\
\ion{Fe}{i} &   613.769  & 2.588 &  1.87 & 4.59 & {1.39} \\
\ion{Fe}{i} &   623.072  & 2.559 &  2.51 & 4.54 & {1.57} \\
\ion{Fe}{i} &   625.256  & 2.404 &  1.64 & 4.56 & {1.50} \\
\ion{Fe}{i} &   639.360  & 2.433 &  2.53 & 4.51 & {1.67} \\
\ion{Fe}{i} &   649.498  & 2.404 &  3.54 & 4.39 & {2.08} \\
\ion{Fe}{i} &   667.799  & 2.692 &  1.58 & 4.62 & {1.29} \\
\ion{Ca}{i} &   671.768  & 2.709 &  0.78 & 4.83 & {0.53} \\
\hline\noalign{\smallskip}
\ion{Li}{i} &   670.784  & 0.000 &  2.37 & 4.97 & {0.00} \\
\hline\noalign{\smallskip}
\end{tabular}
\end{center}
\end{table}

\begin{table*}[htb]
  \vspace*{-2mm}
  \begin{center}
  \caption{List of observed Li\,I $\lambda\,670.8$~nm spectra of eight 
           metal-poor turn-off stars, analyzed in this work for the
           presence of $^6$Li.}
  \label{tab4}
 {\small
  \begin{tabular}{lcccrrcc}
\hline\noalign{\smallskip}
{Star} & {$T_{\rm eff}$~[K]} & {$\log g$} & {[Fe/H]} & 
{R=$\lambda/\Delta\lambda$} &
{S/N$^{~\ast)}$} & {Instrument} & {Reference} \\
\noalign{\smallskip}\hline\noalign{\smallskip}
HD\,74000  & $6203$ & $4.03$ & -2.05 & $120\,000$ & $600^{~~~}$ &
{\sf ESO3.6 / HARPS} & {C07$^{~1)}$} \\
G271$-$162 & $6330$ & $4.00$ & -2.25 & $110\,000$ & $600^{~~~}$ &
{\sf VLT / UVES}     & {N00$^{~2)}$}\\
HD\,84937  & $6310$ & $4.10$ & -2.40 & $100\,000$ & $650^{~~~}$ &
{\sf CFHT / GECKO}   & {C99$^{~3)}$} \\
G020$-$024 & $6247$ & $3.98$ & -1.89 & $120\,000$ & $420^{~~~}$ &
{\sf VLT / UVES}     & {A06$^{~4)}$} \\
HD\,140283 & $5753$ & $3.70$ & -2.40 & $95\,000$ & $1000^{~~~}$ &
{\sf SUBARU / HDS}   & {A04$^{~5)}$} \\
HD\,160617 & $5990$ & $3.79$ & -1.76 & $100\,000$ & $400^{~~~}$ &
{\sf ESO3.6 / HARPS} & {M12$^{~6)}$} \\
G64$-$12   & $6371$ & $4.26$ & -3.24 & $ 95\,000$ & $620^{~~~}$ &
{\sf Keck / HIRES}   & {A08$^{~7)}$}  \\
G275$-$4   & $6338$ & $4.32$ & -3.21 & $ 93\,000$ & $700^{~~~}$ &
{\sf VLT / UVES}     & {N09$^{~8)}$} \\
\hline
  \end{tabular}
  }
 \end{center}
\vspace{1mm}
 \scriptsize{
 {\it Notes:}
  $^{~1)}$ \cite{Cayrel2007}, 
  $^{~2)}$ \cite{Nissen2000},
  $^{~3)}$ \cite{Cayrel1999}, 
  $^{~4)}$ \cite{A2006},
  $^{~5)}$ \cite{Aoki2004}, \\
  $^{~6)}$ L. Monaco and G. Lo Curto (2012, priv. comm.), 
  $^{~7)}$ \cite{A2008}, 
  $^{~8)}$ P.E. Nissen (2009, priv. comm.), \\
  $^{\ast)}$ S/N estimated from the (re-reduced) spectra actually used in 
            this work, may differ from literature values.
 }
\end{table*}

We have simulated this procedure using synthetic lines only. Our
selection of $10$ calibration lines is compiled in Tab.\,\ref{tab3}.
The lines have been selected to have similar wavelength, equivalent
width, and excitation energy (relative to the ionization continuum)
as \ion{Li}{i} $670.8$~nm. Synthetic line profiles generated from 
3D model d3t63g40mm20n01 (assuming LTE) represent the observed line 
profiles and are fitted with 1D LTE profiles from the corresponding 
LHD model to derive $V_{\rm BR}$. The results are given in col.\ (5)
of Tab.\,\ref{tab3}. These numbers have to be compared with
the value of $V_{\rm BR}$ obtained by fitting the 3D non-LTE Li profile
of the same 3D model with 1D LTE profiles from LHD models, 
$V_{\rm BR}=4.97$~km/s (last row of Tab.\,\ref{tab3}).
When fixing $V_{\rm BR}$ to the values implied by the individual
calibration lines, the resulting $^6$Li/$^7$Li isotopic ratio changes 
by $\delta q^{(1D)}$(Li) with respect to the value obtained when
treating $V_{\rm BR}$ as a free fitting parameter. 

We find that, even under these idealized conditions, $\delta q^{(1D)}$(Li) 
is not constant but varies from line to line (see col.\ (6) of 
Tab.\,\ref{tab3}). This is an indication that 1D micro / macro model 
is not perfectly adequate to describe the 3D hydrodynamical velocity 
field. Moreover, since $\delta q^{(1D)}$(Li) is positive for all of 
our calibration lines, the use of calibration lines
leads to an additional overestimation of the Li isotopic ratio
by up to $2$\%. The total error of the 1D analysis is obtained by 
adding the corrections from Tables \ref{tab1} and \ref{tab3}, 
$\Delta q$(Li) = $\Delta q_{(B)}$(Li) + $\delta q^{(1D)}$(Li).

In a further experiment, we have fitted both the calibration lines and 
the Li feature with 3D LTE line profiles. By construction, the derived
$V_{\rm BR}$ is then equal to $FWHM$ of the assumed instrumental profile, 
and is identical for all calibration lines. The Li isotopic ratio
obtained from fitting the 3D non-LTE Li feature with 3D LTE profiles
and fixed $V_{\rm BR}=FWHM$ is $\delta q^{(3D)}$(Li)=$0.054$, compared
to $0.003$ if $V_{\rm BR}$ is a free fitting parameter. The basic reason 
for this alarming result is that the half width of the Li line is 
significantly smaller in LTE than in non-LTE, which must be compensated 
by a higher $^6$Li content. This effect may explain why A06 find higher
$^6$Li/$^7$Li isotopic ratios in their 3D LTE analysis 
(with respect to 1D LTE, their Table 5).

We note that the above results are consistent with previous findings 
by \cite{steffen2010b}, who analyzed the spectrum of HD\,74000 
with a subset of 6 clean \ion{Fe}{i} calibration lines, and obtained
significantly higher $^6$Li/$^7$Li isotopic ratios compared to
the analysis without calibration lines. The same behavior 
is seen by L12, if they fit both the calibration lines and Li with
3D LTE profiles. Using instead non-LTE profiles for fitting both 
\ion{Ca}{i} and \ion{Li}{i} lines leads to consistent $^6$Li/$^7$Li 
isotopic ratios for HD\,84937.

\begin{figure*}[ht]
\vspace*{-4mm}
\mbox{\includegraphics[width=0.8\hsize,bb=40 40 580 370,clip=true]
{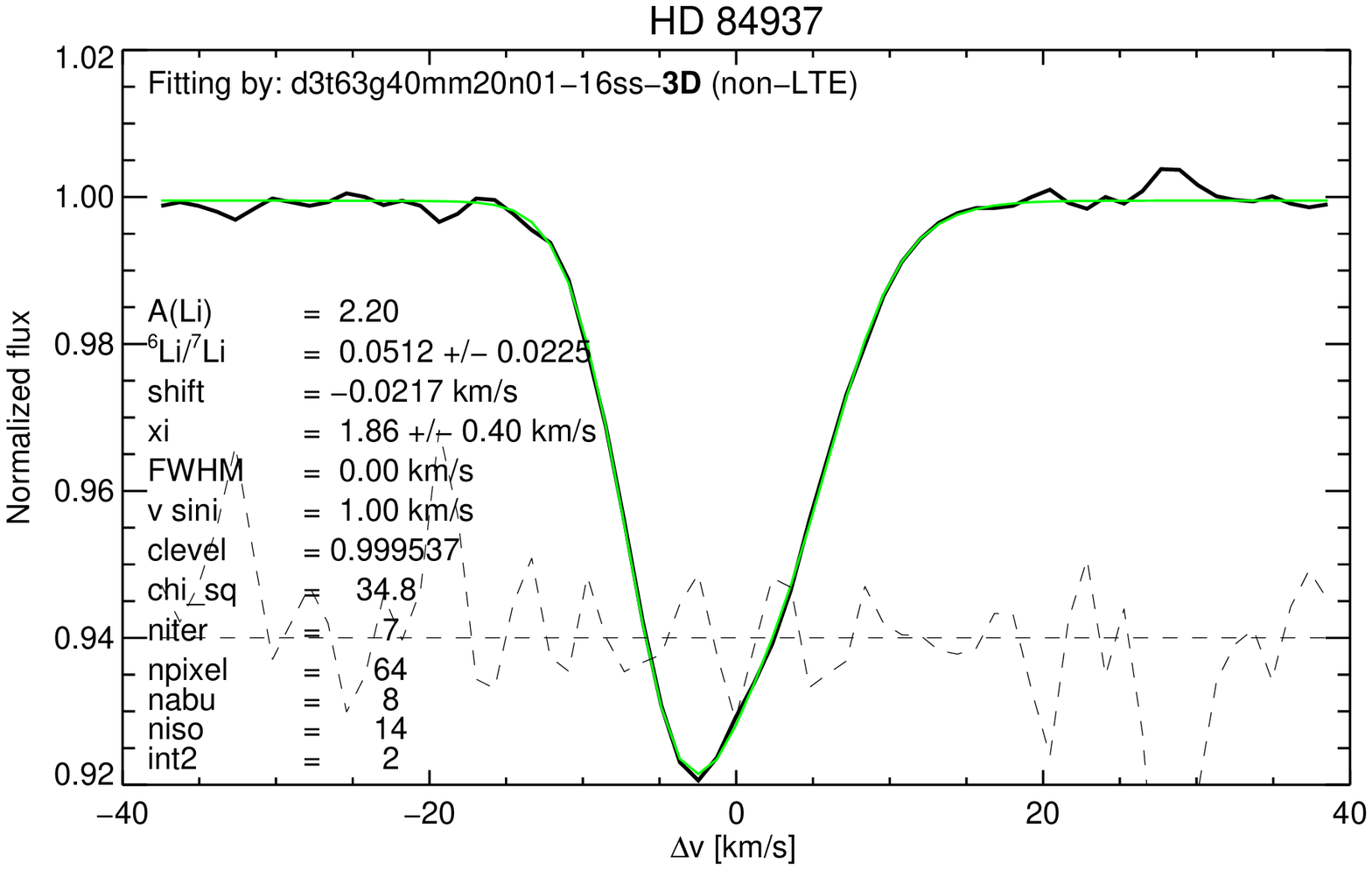}}\\[2mm]
\mbox{\includegraphics[width=0.8\hsize,bb=40 40 580 370,clip=true]
{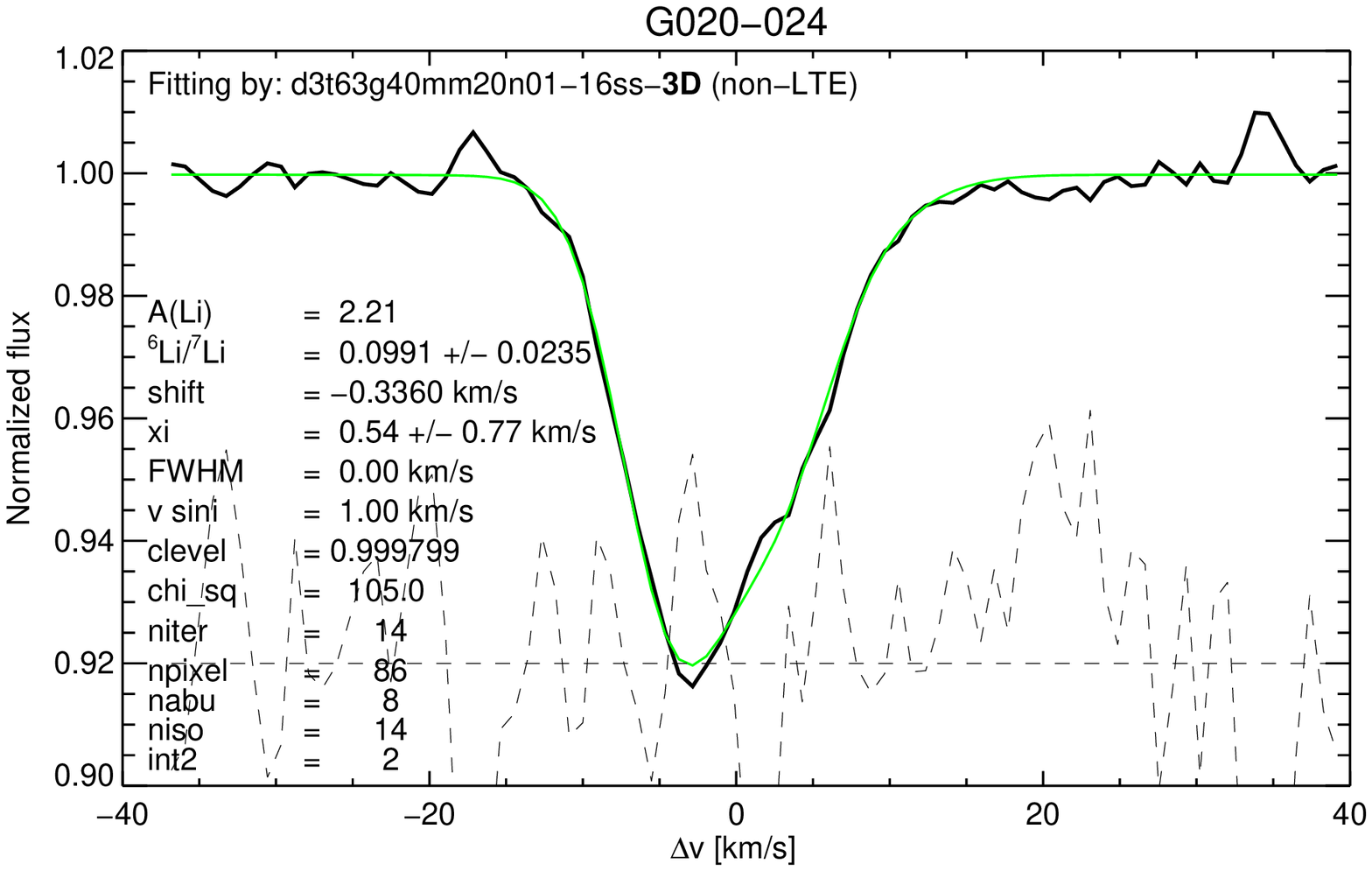}}
\caption{\footnotesize
The best fitting 3D non-LTE Li 670.8~nm line profile (green) superimposed 
on the observed spectrum (solid black) of HD\,84937 (top) and
G020--024 (bottom). The difference between computed and observed 
spectrum ($\times 10$; dashed line) indicates the quality of the fit,
which is much better for HD\,84937 than for G020--024, the latter
suffering from a 'perturbation' in its lower red wing.
}
\label{fig04}
\end{figure*}

\section{Analysis of observed spectra: 3D non-LTE versus 1D LTE line fitting}
\label{s:8stars}
With the results given in Sect.\,\ref{s:3Dbias}, the recommended method of
analysis of the \ion{Li}{i} $670.8$\,nm doublet in an observed spectrum
is as follows. The first step is to produce a grid of 1D LTE synthetic line
profiles as a function of $A$(Li), $q$(Li), using any standard 1D 
mixing-length model with the correct stellar parameters ($T_{\rm eff}$, 
$\log g$, [Fe/H]). This grid is then used to find the best fit to the observed
line profile, as described above (no calibration lines). Finally, the 
resulting $q$(Li) is corrected for 3D effects by subtracting the differential 
correction $\Delta q_{(B)}$(Li) defined in Sect.\,\ref{s:3Dbias}, interpolated 
to the actual stellar parameters according to Eq.\,(\ref{eqn1}) and Table\,\ref{tab2}.

In the following, however, we disregard this differential approach, 
and present instead preliminary results of fitting the observed 
\ion{Li}{i} $\lambda\,670.8$~nm spectra of eight halo turn-off stars 
(see Table\,\ref{tab4}) directly with our grids of 3D~non-LTE and 1D~LTE 
synthetic line profiles, computed from 3D CO$^5$BOLD models and 
1D LHD models, respectively. As described above, five free fitting 
parameters ($A$(Li), $q$(Li), $V_{\rm BR}$, $\Delta v$, $S_{\rm c}$) are 
varied independently to find the best fit. Formally, we 
fix the rotational velocity to $v\,\sin i$=$1.0$~km/s, noting that the 
derived $^6$Li/$^7$Li isotopic ratio is insensitive to this assumption
\citep[see][]{steffen2010b}.

\begin{table*}
  \vspace*{-2mm}
  \begin{center}
  \caption{Fitting the observed spectra of Table\,\ref{tab4}
           with 3D non-LTE and 1D LTE line profiles.}
  \label{tab5}
 {\footnotesize\
  \begin{tabular}{lcccccc}\hline\noalign{\smallskip}
{Star}            & {Model} & {synthetic} & {$A$(Li)}$^{~1)}$ &
{$V_{\rm BR}^{~2)}$} &
 \multicolumn{2}{c}{$q$(Li)\,\,\, [$\times 100$]} \\
{} & {name}  & {spectrum}  & {} & {[km/s]} &
this work & A06,A08,L12 \\
\hline\noalign{\smallskip}
HD\,74000  & \scriptsize d3t63g40mm20n01       & 3D  NLTE & $2.25$ & $2.9 \pm 0.4$ & {$  -0.8 \pm 1.2$} & ---            \\
           & \scriptsize t6280g40mm20a05ob6    & 1D~~~LTE & $2.23$ & $5.7 \pm 0.2$ & {$~~~0.6 \pm 1.1$} & ---            \\
\hline\noalign{\smallskip}                                            
G271$-$162 & \scriptsize d3t63g40mm20n01       & 3D  NLTE & $2.29$ & $3.5 \pm 0.3$ & {$~~~0.6 \pm 1.1$} & ---            \\
           & \scriptsize t6280g40mm20a05ob6    & 1D~~~LTE & $2.27$ & $6.1 \pm 0.2$ & {$~~~2.1 \pm 1.2$} & $~1.9 \pm 1.2$$^{~3)}$   \\
\hline\noalign{\smallskip}                                            
HD\,84937  & \scriptsize d3t63g40mm20n01       & 3D  NLTE & $2.20$ & $3.1 \pm 0.7$ & {$~~~5.1 \pm 2.3$} & $~0.7 \pm 0.5$$^{~5)}$   \\
           & \scriptsize t6280g40mm20a05ob6    & 1D~~~LTE & $2.18$ & $5.8 \pm 0.4$ & {$~~~6.6 \pm 2.4$} & ---            \\
\hline\noalign{\smallskip}                                            
G020$-$024 & \scriptsize d3t63g40mm20n01       & 3D  NLTE & $2.21$ & $0.9 \pm 1.3$ & {$~~~9.9 \pm 2.4$} & ---            \\
           & \scriptsize t6280g40mm20a05ob6    & 1D~~~LTE & $2.19$ & $5.0 \pm 0.4$ & {$~~11.7 \pm 2.5$} & $~7.0 \pm 1.7$$^{~3)}$   \\
\hline\noalign{\smallskip}                                            
HD\,140283 & \scriptsize d3t57g37mm20n01       & 3D  NLTE & $2.20$ & $3.7 \pm 0.1$ & {$-0.6   \pm 0.4$} & $-0.4 \pm 0.2$$^{~5)}$   \\
           & \scriptsize d3t57g37mm30n01       & 3D  NLTE & $2.22$ & $3.3 \pm 0.1$ & {$-1.0   \pm 0.4$} & ---            \\
           & \scriptsize t5775g37mm20ml3a05ob6 & 1D~~~LTE & $2.18$ & $5.5 \pm 0.1$ & {$~~~0.5 \pm 0.4$} & $~0.8 \pm 0.6$$^{~3)}$   \\
           & \scriptsize t5789g37mm30ml3a05ob6 & 1D~~~LTE & $2.19$ & $5.5 \pm 0.1$ & {$~~~0.4 \pm 0.4$} & ---            \\
\hline\noalign{\smallskip}                                            
HD\,160617 &  \scriptsize d3t59g40mm20n02      & 3D  NLTE & $2.14$ & $4.5 \pm 0.2$ & {$  -0.5 \pm 1.2$} & ---            \\
           &  \scriptsize d3t59g35mm20n01      & 3D  NLTE & $2.17$ & $3.3 \pm 0.3$ & {$  -1.3 \pm 1.2$} & ---            \\
           &  \scriptsize t5860g40mm20ml3a05   & 1D~~~LTE & $2.11$ & $5.8 \pm 0.2$ & {$~~~0.1 \pm 1.2$} & $~3.6 \pm 1.0$$^{~3)}$   \\
           &  \scriptsize t5860g35mm20a05      & 1D~~~LTE & $2.13$ & $5.8 \pm 0.2$ & {$~~~0.3 \pm 1.2$} & ---            \\
\hline\noalign{\smallskip}                                            
G64$-$12   & \scriptsize d3t63g40mm30n01       & 3D  NLTE & $2.19$ & $5.0 \pm 0.5$ & {$~~~0.8 \pm 2.7$} & $~2.1 \pm 1.2$$^{~5)}$   \\
           & \scriptsize d3t63g45mm30n01       & 3D  NLTE & $2.19$ & $5.5 \pm 0.5$ & {$~~~1.2 \pm 2.8$} & ---            \\
           & \scriptsize t6270g40mm30ml3a05    & 1D~~~LTE & $2.16$ & $7.1 \pm 0.4$ & {$~~~2.3 \pm 2.9$} & $~5.9 \pm 2.1$$^{~4)}$   \\
           & \scriptsize t6270g45mm30ml3a05ob6 & 1D~~~LTE & $2.14$ & $7.0 \pm 0.4$ & {$~~~2.0 \pm 2.9$} & ---            \\
\hline\noalign{\smallskip}                                            
G275$-$4   & \scriptsize d3t63g40mm30n01       & 3D  NLTE & $2.11$ & $3.6 \pm 0.6$ & {$~~~3.5 \pm 2.4$} & ---            \\
           & \scriptsize d3t63g45mm30n01       & 3D  NLTE & $2.11$ & $4.3 \pm 0.5$ & {$~~~3.8 \pm 2.4$} & ---            \\
           & \scriptsize t6270g40mm30ml3a05    & 1D~~~LTE & $2.08$ & $6.3 \pm 0.4$ & {$~~~4.7 \pm 2.5$} & ---            \\
           & \scriptsize t6270g45mm30ml3a05ob6 & 1D~~~LTE & $2.06$ & $6.2 \pm 0.4$ & {$~~~4.4 \pm 2.4$} & ---            \\
\hline
  \end{tabular}
  }
 \end{center}
\vspace{1mm}
 \scriptsize{
 {\it Notes:}
  $^{1)}$ $\log \left[n(^6{\rm Li})+n(^7{\rm Li})\right] -
  \log n({\rm H)}+12$; $^{2)}$ Gaussian kernel,
  $^{~3)}$ \cite{A2006}, $^{~4)}$ \cite{A2008}, \\
  \hspace*{6mm}$^{~5)}$ \cite{L2012}
 }
\end{table*}

The results are presented in Table\,\ref{tab5}. 
The errors quoted for $V_{\rm BR}$ and $q$(Li) 
are the formal $1\,\sigma$ confidence intervals due to the finite 
S/N of the fitted spectra. They are computed in the usual way 
\citep[see e.g.][]{Nissen2000,A2006} by finding 
the distance $\Delta p$ of the parameter of interest $p$ from is optimum 
value $p_0$ such that $\chi^2(p+\Delta p) - \chi^2(p_0)\equiv 
\Delta \chi^2 =1$, fixing $p$ and minimizing $\Delta \chi^2$ over
the remaining free fittting parameters. 
Here $\chi^2$ is defined as 
\begin{equation}
\label{eqn2}
\chi^2 = \sum_i\,\frac{( O_i-S_i)^2}{\sigma^2}\, ,
\end{equation}
where $O_i$ and $S_i$ are the observed and synthetic flux at
wavelength point $i$, and $\sigma$=(S/N)$^{-1}$, with S/N taken
from Tab.\,\ref{tab4}. Note that this is a lower limit to the real error,
which may have several other significant contributions 
\citep[see discussion in][]{GarciaPerez2009}.

As expected, the 3D analysis yields systematically lower $^6$Li/$^7$Li
isotopic ratios by up to 1.8\%. For comparison, we list in the last
column the 1D LTE results of A06, A08, and the 3D non-LTE results of
L12 from their case (a) for the stars in common with our 
sample. In some cases, the agreement is very good (G271--162, HD140283), 
even though the analysis is based on different observational data. In 
other cases, we obtain significantly lower $^6$Li/$^7$Li ratios 
(HD\,160617, G64--12). The most striking discrepancy, this time in the 
opposite direction, is seen for HD\,84937.

G020--024 is a special case. We have retrieved the spectra obtained 
by A06 from the ESO archive and re-reduced them with the
current UVES pipeline. We note that all six sub-exposures show an
unexplained 'perturbation' in the lower red wing of the Li $670.8$~nm 
line that cannot be fitted properly, neither with 1D LTE nor with
3D non-LTE model spectra. The best 3D non-LTE fit suggests a very high 
$^6$Li content near $10\%$ (see Fig.\,\ref{fig04}), even exceeding 
the 1D LTE result by A06 of $7\%$. In view of the poor quality of
the fit, we consider this result as doubtful. It is worth mentioning
that G020--024 is listed as a suspected binary in \cite{Fouts1987};
moreover, A06 report that this star shows an unusually large 
discrepancy between the photometric and the H$_\alpha$-based $T_{\rm eff}$.
Possibly, G020--024 is a spectroscopic binary whose components are of 
similar spectral type.

Disregarding G020--024, we are left with one formal
$2\,\sigma$ detection (HD\,84937), one $1\,\sigma$ 
detection (G275--4), and five non-detections (HD\,74000,
G271--162, HD\,140283, HD160617, G64--12) out of the 
eight stars, when considering the 3D non-LTE results only.
In 1D LTE, G271--162 would turn into a formal $1\,\sigma$ 
detection (see Tab.\,\ref{tab5}).

\section{Conclusions}
The $^6$Li/$^7$Li isotopic ratio derived by fitting the Li\,I 
doublet with 3D non-LTE synthetic line profiles is shown to be
1\% to 2\% lower than what is obtained with 1D LTE profiles. 
This result implies that only $2$ out of the
$24$ stars of the \cite{A2006} sample would formally remain 
significant ($3\,\sigma$) $^6$Li detections when subjected to a 
3D~non-LTE analysis.

In another theoretical case study we have demonstrated that the
difference between 3D non-LTE and 1D LTE results increases even 
more if we would rely on additional 'calibration lines' to 
fix the residual line broadening, as advocated e.g.\ by A06, A08, L12.
The number of possible $^6$Li detections is thus further reduced.
We conclude that the usage of additional 'calibration lines', even
if carefully selected, introduces additional uncertainties rather 
than reducing the error of the analysis, unless the 3D non-LTE line
formation is fully understood for all involved lines.

Finally, we have analyzed available high quality spectra of eight 
turn-off halo stars, both with 1D LTE and 3D non-LTE modeling. In 
most cases, the evidence for the presence of $^6$Li is not significant. 
Only in the case of HD\,84937 it is difficult to deny the signature 
of $^6$Li. Surprisingly, L12 no longer find any significant evidence
for the presence of $^6$Li in this object, which since almost two
decades has been considered as an undisputed $^6$Li detection. From
the results reported by L12, it is unclear whether the reason for this
disturbing discrepancy is related to the new superior observational 
data or to the improved method of analysis.

We conclude that 3D model atmospheres can indeed help to solve the
\emph{second lithium problem}. In view of the 3D non-LTE results reported 
in this work (and by L12), it seems that the presence of $^6$Li in the
atmospheres of galactic halo stars is rather the exception than the rule, 
and hence does not necessarily constitute a \emph{cosmological} problem.

\begin{acknowledgements}
We thank L. Monaco and G. Lo Curto for re-reducing the HARPS 
spectrum of HD\,160617, P.E. Nissen for providing us with a 
carefully reduced UVES spectrum of G275-4, and W. Aoki for
allowing us to use his HDS spectrum of HD\,140283.
\end{acknowledgements}

\bibliographystyle{aa}

\begin{thebibliography}{}

\bibitem[{Aoki \etal (2004)}]{Aoki2004} 
Aoki, W., Inoue, S., Kawanomoto, S., \etal 2004, 
\aap, 428, 579 

\bibitem[{Asplund \etal (2006)}]{A2006} 
Asplund, M., Lambert, D.L., Nissen, P.E., Primas, F., 
Smith, V.V.\ 2006, (A06) \apj, 644, 229 

\bibitem[{Asplund \& Melendez(2008)}]{A2008} 
Asplund, M., Mel{\'e}ndez, J.\ 2008, (A08)
\textit{First Stars III}, 990, 342 

\bibitem[Barklem, \etal (2003)]{barklem2003}
Barklem, P.S., Belyaev, A.K., Asplund, M.\ 2003, \textit{A\&A}, 409, L1

\bibitem[Castelli \& Kurucz (2004)]{CK2004}
{Castelli, F., \& Kurucz, R.~L.} 2004, arXiv:astro-ph/0405087

\bibitem[Cayrel \etal\ (1999)]{Cayrel1999}
{Cayrel, R., Spite, M., Spite, F., Vangioni-Flam, E., Cass\'e, M.,
 Audouze, J.} 1999, \aap, 343, 923

\bibitem[{Cayrel \etal (2007)}]{Cayrel2007} 
Cayrel, R., Steffen, M., Chand, H., \etal\ 2007, \aap, 473, L37

\bibitem[Cayrel \etal (2008)]{Cayrel2008} {Cayrel, R.,
Steffen, M., Bonifacio, P., Ludwig, H.-G., Caffau, E.} 2008,
\textit{Nuclei in the Cosmos (NIC~X)}, availale online at  
\url{http://adsabs.harvard.edu/abs/2008nuco.confE...2C}

\bibitem[Christlieb(2008)]{Ch2008} {Christlieb, N.} 2008,
\textit{Journal of Physics G: Nuclear Physics}, 35, 014001

\bibitem[Fouts(1987)]{Fouts1987} Fouts, G.\ 1987, \pasp, 99, 986 

\bibitem[Freytag \etal (2002)]{F2002}
Freytag, B., Steffen, M., \& Dorch, B.\ 2002, 
\textit{AN}, 323, 213

\bibitem[Freytag \etal (2012)]{F2012}
 Freytag, B., Steffen, M., Ludwig, H.-G., \etal\ 2012, 
\textit{Journal of Computational Physics}, 231, 919 

\bibitem[Garc{\'{\i}}a P{\'e}rez \etal (2009)]{GarciaPerez2009} 
Garc{\'{\i}}a P{\'e}rez, A.E., Aoki, W., Inoue, S., \etal\ 2009, 
\aap, 504, 213 

\bibitem[Lind \etal (2012)]{L2012} Lind, K., Asplund, M., 
Collet, R., Mel\'endez, J.\ 2012 (L12),
\textit{\mbox{MemSAI}}, (this volume)

\bibitem[Ludwig \etal (2009)]{L2009} Ludwig, H.-G., Caffau, E.,
{Steffen, M., Freytag, B., Bonifacio, P., \& Ku\v{c}inskas, A.} 2009,
\textit{\mbox{MemSAI}}, 80, 708


\bibitem[Nissen \etal (2000)]{Nissen2000}
{Nissen, P.E., Asplund, M., Hill, V., D'Odorico, S.}
2000, \aap 357, L49



\bibitem[Prantzos(2010)]{Prantzos2010} {Prantzos, N.} 2010,
\textit{IAU Symposium, 268}, 473 

\bibitem[Prantzos(2012)]{Prantzos2012} {Prantzos, N.} 2012,
\textit{arXiv:1203.5662} (\aap, in press)

\bibitem[Sbordone \etal (2010)]{S2010}
{Sbordone, L., Bonifacio, P., Caffau, E., \etal}
 2010, \aap, 522, A26 

\bibitem[Steffen \etal (2010a)]{steffen2010a}
{Steffen, M., Cayrel, R., Boinfacio, P., Ludwig, H.-G., Caffau, E.}
2010a,  IAU Symposium, 265, 23 

\bibitem[Steffen \etal (2010b)]{steffen2010b}
{Steffen, M., Cayrel, R., Boinfacio, P., Ludwig, H.-G., Caffau, E.}
2010b, IAU Symposium, 268, 215 

\end{thebibliography}

\end{document}